\documentclass[fleqn,twoside,twocolumn,nofootinbib,showkeys]{revtex4} 
\usepackage[sec,nocpr]{ujp} 
\newcommand{\be}{\begin{equation}}
\newcommand{\ee}{\end{equation}}
\newcommand{\bee}{\begin{eqnarray}}
\newcommand{\eee}{\end{eqnarray}}
\usepackage{color}
\usepackage{url}

\begin{document}
\title[Nucleon momentum distributions in $^3$He]%
{Nucleon momentum distributions in $^3$He and three-body interactions }%
\author{S.V.~Bekh}
\affiliation{National Technical University of Ukraine ``Igor Sikorsky Kiev Polytechnic Institute''}
\address{
37, Prospect Peremogy, 03056 Kiev, Ukraine}
\email{bekh@bitp.kiev.ua}
\author{A.P.~Kobushkin}
\affiliation{Bogolyubov Institute for Theoretical Physics, Nat. Acad. of Sci. of Ukraine}
\address{14b, Metrolohichna Str., Kiev 03143, Ukraine}
\email{kobushkin@bitp.kiev.ua}
\affiliation{National Technical University of Ukraine ``Igor Sikorsky Kiev Polytechnic Institute''}
\address{Prospect Peremogy 37, 03056 Kiev, Ukraine}
\author{E.A.~Strokovsky}
\affiliation{Laboratory of High Energy Physics, Joint Institute for Nuclear Research }
\address{141980, Dubna, Russia}
\email{ strok@jinr.ru}
\affiliation{ Research Center for Nuclear Physics, Osaka University}
\address{10-1 Mihogaoka, Ibaraki, Osaka, 567-0047, Japan}

\udk{539.1.01, 539.141, 539.17}
\razd{Nuclear reactions}

\autorcol{S.V.~Bekh, A.P.~Kobushkin, and E.A.~Strokovsky}%

\setcounter{page}{1}%

\begin{abstract}
We calculate the momentum distributions of neutrons and protons in $^3\mathrm{He}$ 
in the framework
of a model which includes 3N interactions together with 2N interactions. It is shown
that the
contribution of 3N interactions becomes essential in comparison with that
coming from 2N interactions for the internal momentum in  $^3\mathrm{He}$
$k>$~250~MeV/$c$. We also compare the calculated momentum distribution of protons with
the so-called empirical momentum distribution of protons extracted from the
$A\mathrm (^3\mathrm{He},p)$ breakup cross-sections
measured for protons emitted at zero degree. It is concluded that
3N interactions
cannot completely explain the disagreement between the available data on the empirical
momentum distribution of protons in $^3\mathrm{He}$ and
calculations based on 2N interactions, which is observed at the
high momentum region of the momentum distribution,  $k>$~250~MeV/$c$.
\end{abstract}


\maketitle

\section{Introduction \label{sec:1}}
Momentum distributions of nucleons in nuclei are directly connected with the
spatial structure of the corresponding nuclear systems. In particular, these
distributions at Fermi momenta above 200-300 MeV/$c$ (this region is usually referred
to as ``a region of high relative nucleon momenta'')
give important information about such interesting  questions as
a role of non-nucleon degrees of freedom in the nuclear structure, relativistic
effects, and so on.

Starting from three nucleon systems, $^3$He and $^3$H, the momentum distributions
should also give information about the role of effective three-nucleon (3N) interactions
in nuclear structure.
For example, a prominent role of 3N interactions was demonstrated in a
systematic study of the elastic scattering of polarized
protons from deuterons at energies from 100 to 200 MeV \cite{Ermisch}. Besides
that, the relativistic effects are also important in 3N systems, see, e.g.,
the results of recent relativistic calculations of the triton biding
energy \cite{Kamada} with the so-called Kharkov potential, one-boson-exchange
NN potential constructed with use of
an unitary clothing transformation \cite{DuSh}.

The goals of this paper are:
\begin{enumerate}
 \item to find signals of manifestation of 3N interactions in the momentum distributions of neutrons and protons in $^3$He,
 \item to compare theoretical results, coming from
 known models for 2N+3N interactions, with existing experimental data,
 \item to indicate what region of relative nuclear momenta  should be looked for manifestations
of non-nucleonic degrees of freedom in the nuclear structure.
\end{enumerate}

The paper is organized in the following way. We start, in Section~\ref{sec:2},
with a short overview of the operator form of a three nucleon bound state, which is a
basic point for further calculations.  In Section~\ref{neutron},
 the momentum distribution of neutrons in $^3$He is calculated within a model, which
takes into account 3N interactions together with the standard 2N interactions.

In Section~\ref{proton}, the momentum distribution of protons in $^3$He
is calculated in the framework of a similar model.
The calculated proton momentum distribution is compared with existing experimental
data in Section~\ref{sect:EMD}, namely: in Subsection~\ref{sect:EMD}.1,
we discuss the definition and a procedure of extraction of the
so-called ``empirical momentum distribution'' of protons in $^3$He from the
A$(^3\mathrm{He},p)$ breakup cross-sections \cite{ALPHA},
when the proton-spectator was emitted at $0^{\circ}$; in Subsection~\ref{sect:EMD}.2,
the empirical momentum distribution is compared with the results of our calculations,
as well as with calculations without explicit inclusion of 3N interactions.
Conclusions are given in Section~\ref{sec:concl}.
\section{Operator form of three nucleon bound state\label{sec:2} }
There are few known approaches to describe a three nucleon
(3N) wave function: a partial wave
decomposition (see, e.g., Ref.~\cite{Baru_et_al}), tensor representations
\cite{KotlyarShebeko, KotlyarShcheglova, KotlyarJourdan}, and an operator
form \cite{GerjouySchwinger}. In this paper we use the last one.

In 1942 E.~Gerjuoy and J.~Schwinger introduced an operator form for
three- and four-nucleon states~\cite{GerjouySchwinger}, which was a generalization
of an operator form of the deuteron state elaborated earlier by
W.~Rarita and J.~Schwinger~\cite{RaritaSchwinger}.
In the case of a 3N nucleon state,
this approach expresses the general spin structure of
a 3N system in terms of nine operator forms acting on the special spin state,
where nucleons 1 and 2 have the total spin $s=0$ and nucleon 3 carries out the spin of
the 3N system:
\be\label{1}
\left|\nu \right\rangle =
\frac{1}{\sqrt2}\left( \left|+-\ \mathrm{sign}\, \nu\right\rangle - \left|-+\ \mathrm{sign}\, \nu\right\rangle\right).
\ee
In Eq.~(\ref{1}) $\nu$ is the magnetic quantum number
of the 3N system and
$$ \arrowvert \mathrm{sign}\, m_1\ \mathrm{sign}\, m_2\ \mathrm{sign}\, m_3\rangle $$
is a spin wave function of three nucleons with magnetic quantum numbers $m_1$, $m_2$,
and $m_3$.

The operator form does not employ the isospin formalism,
and the nucleons are labelled as follows:\\[0.5cm]
\begin{tabular}{ll}
$N_1 = N_2 = p$ and $N_3 = n$\hspace{0.25cm} &--- for $^3$He,\\
$N_1 = N_2 = n$ and $N_3 = p$\hspace{0.25cm} &--- for $^3$H.
\end{tabular}
\\[0.5cm]

The relations between approaches, which employ
(or do not employ)
the isospin formalism,
as well as advantages of the latter ones, were discussed in Refs.~\cite{SDG, DS}.

It was mentioned in Ref.~\cite{Fachruddin_et_al},
that the ninth spin structure of the operator form of a
 3N system is redundant and we,
following to Ref.~\cite{Fachruddin_et_al}, omit this component.

Finally, the 3N bound state wave function is given by
 \be\label{2}
 \Psi_\nu(\mathbf p, \mathbf q) = \sum_{i=1}^8 \phi_i(p,q,x) \left|i, \nu \right\rangle ,
 \ee
where $ \left|i, \nu \right\rangle$ are the spin wave functions defined below
(see, Eqs.~(\ref{spin_structures_3-8})), $\mathbf p$ and $\mathbf q$ are the Jacobi momenta
\be\label{Jacobi}
\begin{split}
& \mathbf{p}_1=\tfrac13\mathbf{P} - \tfrac12\mathbf{q} + \mathbf{p}\,,\quad
 \mathbf{p}_2=\tfrac13\mathbf{P} - \tfrac12\mathbf{q} - \mathbf{p}\,,\\
& \mathbf{p}_3=\tfrac13\mathbf{P} + \mathbf{q}\, . 
\end{split}
\ee
 Here, $\mathbf p_1$, $\mathbf p_2$, and $\mathbf p_3$ are the
momenta of the nucleons, and $\mathbf P$ is the
momentum of the nucleus; $x = \cos \kappa$ ($\kappa$ is the
angle between the vectors $\mathbf p$ and $\mathbf q$),
 and $\phi_i(p,q,x)$ are scalar functions.
The scalar functions $\phi_i(p,q,x)$ have been
calculated in Ref.~\cite{Fachruddin_et_al} for two modern potentials:
the 2N potential AV18 \cite{AV18} with the 3N potential Urbana-IX \cite{U9}
(AV18+U9) and the 2N potential CD-Bonn \cite{CDB}  with the 3N
potential Tucson-Melbourne \cite{TM} (CDBN+TM).
The functions $\phi_i(p,q,x)$ are tabulated
on a 3-dimensional grid $(x,q,p)$ and can be downloaded from
site \cite{grid}.

The spin structures are given as follows:
\be\label{spin_structures_3-8}
\begin{split}
\left|1\nu\right\rangle&= \left|\nu\right\rangle,\quad
\left|2\nu\right\rangle = \sqrt{\tfrac13}
\bigl(\mbox{\boldmath $\sigma$}(12)\cdot \mbox{\boldmath $\sigma$}(3)\bigr)\left|\nu\right\rangle,\\
\left|3\nu\right\rangle&=-i\sqrt{\tfrac32}\bigl(\mbox{\boldmath $\sigma$}(3) \cdot
\left(\widehat{\mathbf p} \times \widehat{\mathbf q}\right)\bigr) \left|\nu\right\rangle ,
\\
\left|4\nu\right\rangle&=\sqrt{\tfrac12}\left[i\mbox{\boldmath $\sigma$}(12)
+ \bigl(\mbox{\boldmath $\sigma$}(12)\times \mbox{\boldmath $\sigma$}(3)\bigr)\right]\cdot
\left(\widehat{\mathbf p} \times \widehat{\mathbf q}\right)\\
& \hspace{5.cm} \times\left|\nu\right\rangle,
\\
\left|5\nu\right\rangle&=\left[-i\mbox{\boldmath $\sigma$}(12)
+\tfrac12\bigl(\mbox{\boldmath $\sigma$}(12)\times \mbox{\boldmath $\sigma$}(3)\bigr)\right]\cdot
\left(\widehat{\mathbf p} \times  \widehat{\mathbf q}\right)\\
& \hspace{5.cm} \times\left|\nu\right\rangle,
\\
\left|6\nu\right\rangle&=\sqrt{\tfrac32}
\left[\bigl(\mbox{\boldmath $\sigma$}(12)\cdot\widehat{\mathbf p}\bigr)\,
\bigl(\mbox{\boldmath $\sigma$}(3)\cdot \widehat{\mathbf p}\bigr)
-\tfrac13\bigl(\mbox{\boldmath $\sigma$}(12)\cdot \mbox{\boldmath $\sigma$}(3)\bigr)
\right] \\
& \hspace{5.cm}
\times \left|\nu\right\rangle,
\\
\left|7\nu\right\rangle&=\sqrt{\tfrac32}
\left[\bigl(\mbox{\boldmath $\sigma$}(12) \cdot \widehat{\mathbf q}\bigr) \,
\bigl(\mbox{\boldmath $\sigma$}(3) \cdot  \widehat{\mathbf q}\bigr)
-\tfrac13\bigl(\mbox{\boldmath $\sigma$}(12)\cdot \mbox{\boldmath $\sigma$}(3)\bigr)
\right]\\
& \hspace{5.cm} \times
\left|\nu\right\rangle,
\\
\left|8\nu\right\rangle&=\tfrac3{2\sqrt5}\left[
\bigl(\mbox{\boldmath $\sigma$}(12)\cdot\widehat{\mathbf q} \bigr)\,
\bigl(\mbox{\boldmath $\sigma$}(3)\cdot \widehat{\mathbf p}\bigr)\right. \\
& \left.\hspace{-0.5cm}
+\bigl(\mbox{\boldmath $\sigma$}(12)\cdot \widehat{\mathbf p}\bigr) \,
\bigl(\mbox{\boldmath $\sigma$}(3)\cdot \widehat{\mathbf q}\bigr)
-\tfrac23\bigl(\widehat{\mathbf p} \cdot \widehat{\mathbf q}\bigr) \,
\bigl(\mbox{\boldmath $\sigma$}(12)\cdot \mbox{\boldmath $\sigma$}(3)\bigr)\right]
\left|\nu\right\rangle,
\end{split}
\ee
%
where
$\widehat{\mathbf q} =\mathbf q /|\mathbf q|$, $\widehat{\mathbf p} =\mathbf p /|\mathbf p|$,
and
\be\label{odd_spin_operator}
\mbox{\boldmath $\sigma$}(12)=
\tfrac12 [\mbox{\boldmath $\sigma$}(1)-\mbox{\boldmath $\sigma$}(2)]\, ;
\ee
$\mbox{\boldmath $\sigma$}(i)$  are the Pauli matrices of $i$-th nucleon.

The normalization of the wave function is given by
\bee
&& \int d^3q d^3 p \left|\Psi_\nu(\mathbf p, \mathbf q)\right|^2 \nonumber \\
 &&=8\pi^2\int_{-1}^1 dx\int_0^\infty d q q^2\int_0^\infty dp p^2 \nonumber\\
 && \times
\left[
\sum_{i=1}^8 \langle i\nu \left|i\nu\right\rangle \phi_i^2(p,q,x) \right.\label{normalization}\\
 && +\left.
2\sum_{i \ne j} \langle i\nu \left|j\nu\right\rangle
\phi_i(p,q,x)\phi_j(p,q,x)
\right] = 1.\nonumber
\eee
Overlaps of the spin structures are given in Table~\ref{tab:1}.

Note that the contributions
of $\phi_3(p,q,x)$, $\phi_4(p,q,x)$, and $\phi_5(p,q,x)$
 to the normalization relation~(\ref{normalization})
are of order of $\sim$ 0.05\% and we ignore these components of the trinucleon wave
function in Table~\ref{tab:1}, as well as in the subsequent calculations.

\begin{table}
\caption{\label{tab:1}
Overlaps of the spin structures $\langle i \nu | j \nu \rangle$.
Only nonvanishing overlaps are presented.}
\begin{center}
\begin{tabular}{|ll|l||ll|l|}
\hline
$i$ & $j$ & $\langle i \nu | j \nu \rangle$ & $i$ & $j$ & $\langle i \nu | j \nu \rangle$  \\
\hline
1   & 1   & 1                               & 6   & 8   & $\sqrt{\tfrac65} x$ \\[0.25cm]
2   & 2   & 1                               & 7   & 7   & 1                   \\[0.25cm]
6   & 6   & 1                               & 7   & 8   & $\sqrt{\tfrac65} x$ \\[0.25cm]
6   & 7   & $\tfrac12 (3x^2-1)$             & 8   & 8   & $\tfrac9{10}(1+\tfrac13x^2)$\\[0.25cm]
\hline
\end{tabular}
\end{center}
\end{table}

We have found that the results of
numerical calculations, published in Refs.~\cite{Fachruddin_et_al, grid},
are represented on the 3-dimensional grid $(p,q,x)$ which is not ``dense'' enough for
needs of our calculations.

Therefore we expanded the scalar functions $\phi_i(p,q,x)$ at fixed $p$ and $q$ in
series in terms of the Legendre polynomials $P_\ell(x)$:
\be\label{analytical1}
\begin{split}
 \phi_1(p,q,x) &= C_{10}(p,q) + C_{12}(p,q) P_2(x),\\
 \phi_2(p,q,x) &= C_{21}(p,q) P_1(x) + C_{23}(p,q) P_3(x),\\
 \phi_3(p,q,x) &= C_{31}(p,q) P_1(x),\\
 \phi_4(p,q,x) &= C_{40}(p,q) + C_{42}(p,q) P_2(x),\\
 \phi_5(p,q,x) &= C_{50}(p,q) + C_{52}(p,q) P_2(x),\\
 \phi_6(p,q,x) &= C_{61}(p,q) P_1(x),\\
 \phi_7(p,q,x) &= C_{71}(p,q) P_1(x),\\
 \phi_8(p,q,x) &= C_{80}(p,q) + C_{82}(p,q) P_2(x).\\
\end{split}
\ee
Terms with the Legendre polynomials of higher orders on $\ell$ were found to be
negligibly small and will be omitted in the present numerical calculations.
For example, in case of functions $\phi_3$, $\phi_6$, and $\phi_7$, the numerical coefficients
for the next term, containing $P_3(x)$,
were found approximately in
$10^{-5}$~--~$10^{-6}$ times less then $C_{31}(p,q)$, $C_{61}(p,q)$, and $C_{71}(p,q)$, respectively.

The coefficients $C_{m \ell}(p,q)$
of the series form 13 functions given on a 2-dimensional grid $(p,q)$.
\section{Momentum distribution of neutrons in $^3$He \label{neutron}}
\begin{figure}\includegraphics[
 height=0.375\textheight]{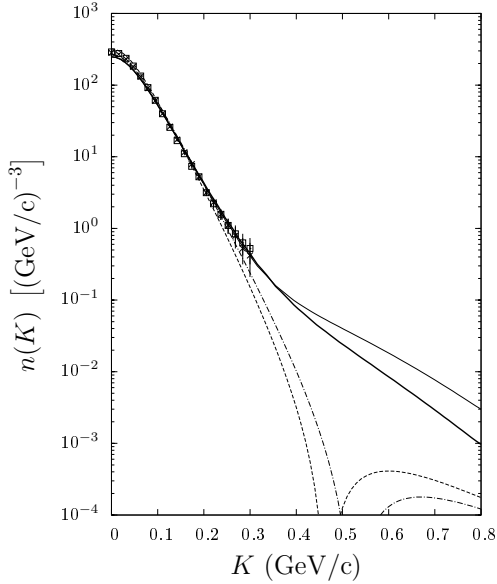}
 \vskip-3mm\caption{Momentum distribution of neutrons in $^3$He calculated with
 2N+3N interactions,
AV18+U9 (thin solid line) and CDBN+TM (thick solid line).
The results obtained with 2N interactions
only (dashed and dot-dashed lines, for the Paris~\cite{Paris} and
CD-Bonn~\cite{CDB} potentials, respectively) are taken from Ref.~\cite{KS}. The
 squares and crosses represent the
 results of variational calculations \cite{SPW} obtained with the Urbana+U9
and Argonne+U9 interactions, respectively.
\label{fig:neutMD}
 }
\end{figure}
\begin{figure}
 \includegraphics[
 height=0.375\textheight]{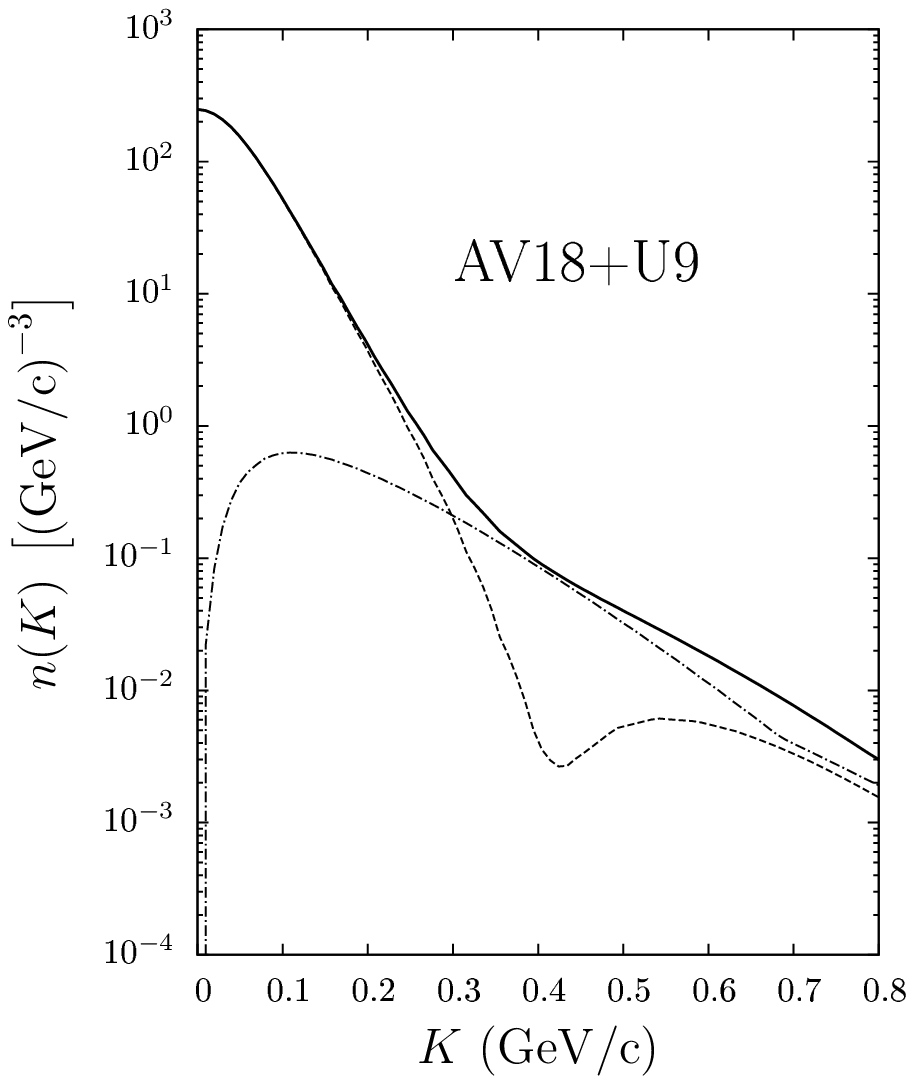}
 \includegraphics[
 height=0.375\textheight]{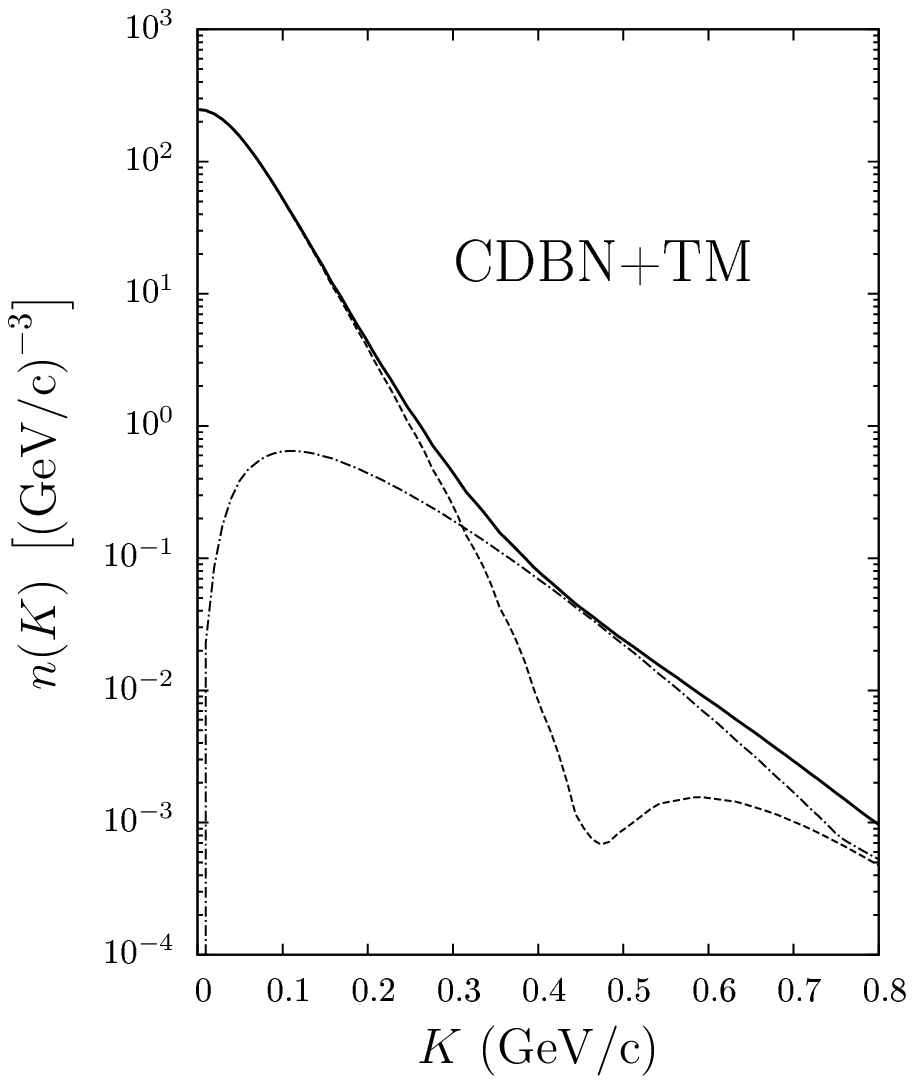}
 \vskip-3mm\caption{
 Contributions
of the main spin structures  to the neutron momentum distribution in $^3$He.
Dashed line: $\left(\phi_1\right)^2$; dot-dashed line: sum of contributions coming
from $\left(\phi_{2,6,7,8}\right)^2$ and interference terms $\phi_6\phi_7$, $\phi_6\phi_8$,  $\phi_7\phi_8$;
solid line: the result with all the terms taken into account.
\label{fig:neutMD_PW}
 }
 \end{figure}

The momentum distribution of a neutron in $^3$He is defined as follows
(see \cite{Fachruddin_et_al}):
\be
\begin{split}
n(K)&=\tfrac12\sum_\nu \int d^3p d^3q \delta(\mathbf q -
\mathbf K)\left|\Psi_\nu (\mathbf p,\mathbf q)\right|^2 \nonumber\\
&= \int d^3p \left|\Psi_\frac12 (\mathbf p,\mathbf K)\right|^2, \label{n(k)_definition}
\end{split}
\ee
$\mathbf K$ is the neutron momentum inside $^3$He.
The factor $\frac12$ comes from averaging over the nucleus magnetic quantum numbers.
Using the spin structures $\langle i \nu | j \nu \rangle$ from Table~\ref{tab:1}, we get
\be\label{n(k)}
 n(K)=2\pi \int_0^\infty p^2 dp \int_{-1}^1 dx\, \rho_n(p,K,x),
\ee
where
\be \label{Phi}
\begin{split}
 \rho_n(p,K,x) &=
 \phi_1^2(p,K,x) +
 \phi_2^2(p,K,x) \\
 & \hspace{-0.95cm}+
 \phi_6^2(p,K,x)  + \phi_7^2(p,K,x) \\
  & \hspace{-0.95cm}+
 \tfrac9{10}\left(1 + \tfrac 13 x^2\right)\phi_8^2(p,K,x)  \\
 &  \hspace{-0.95cm}+
 \left(-1 + 3 x^2\right)\phi_6(p,K,x)\phi_7(p,K,x) \\
 &  \hspace{-0.95cm} +
 \sqrt{\tfrac{24}5}x\left[\phi_6(p,K,x) + \phi_7(p,K,x) \right]\phi_8(p,K,x) .
\end{split}
\ee
The resulting neutron momentum distribution, $n(K)$, for AV18+U9 and CDBN+TM together
with the results of variational calculations from Ref.~\cite{SPW},
 are shown in Fig.~\ref{fig:neutMD}. We compare this result with
calculations from Ref.~\cite{KS} obtained without 3N interactions. Good agreement
between the results, obtained with and without 3N interactions,
 is obvious for $K \lesssim$~250~MeV/$c$ and
demonstrates that 
3N interactions do not manifest itself in this region.
At higher $K$, the contribution of 
3N interactions becomes significant and
dominates from $K \sim$~400~MeV/$c$ over the one from 2N interactions.

The contribution of the most important term, $\left(\phi_1\right)^2$, and
the sum of other terms  in Eq.~(\ref{Phi}) to the total momentum distribution
are displayed in Fig.~\ref{fig:neutMD_PW}. It is worthwhile
to note
that contribution of the $\left(\phi_1\right)^2$ term has a dip in the same region
(near $K\sim$~450 MeV/$c$), where the similar dip appears in calculations without
3N interactions.
\section{Momentum distribution of protons in $^3$He \label{proton}}
\begin{figure}
 \includegraphics[
 height=0.375\textheight]{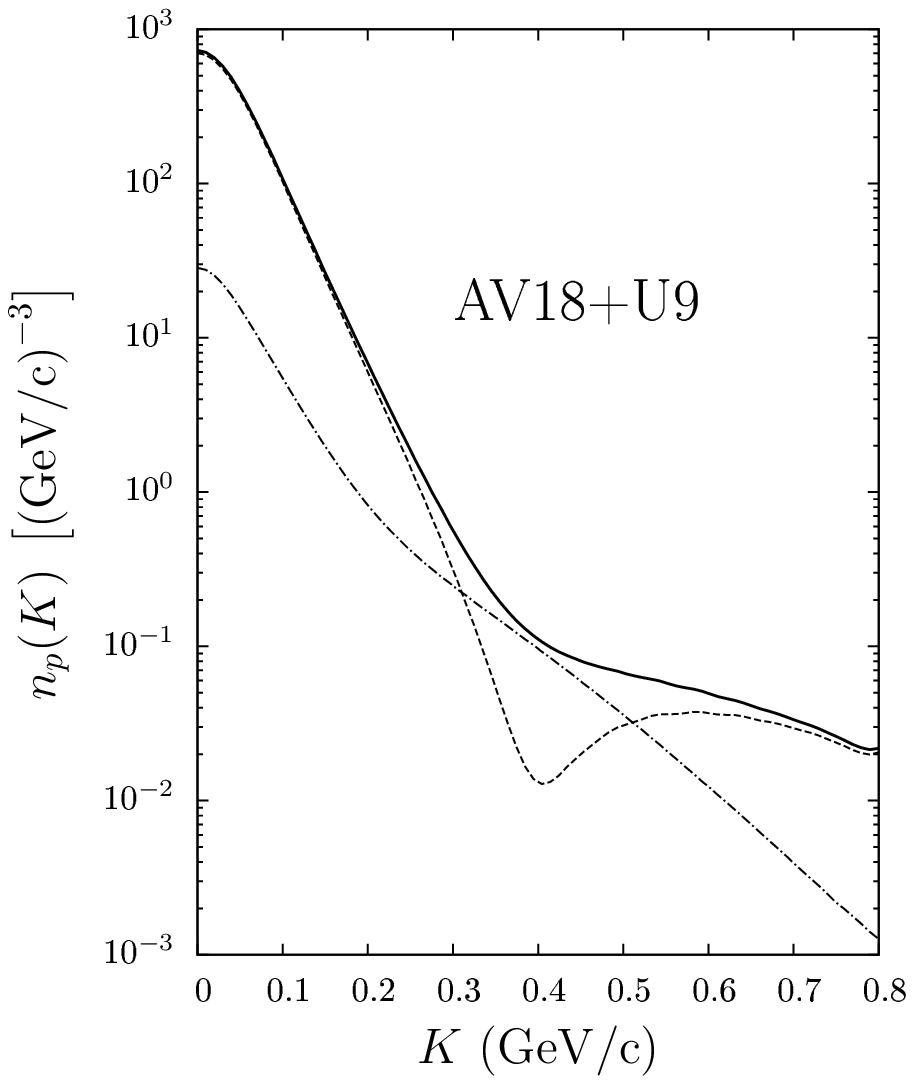}
 \includegraphics[
 height=0.375\textheight]{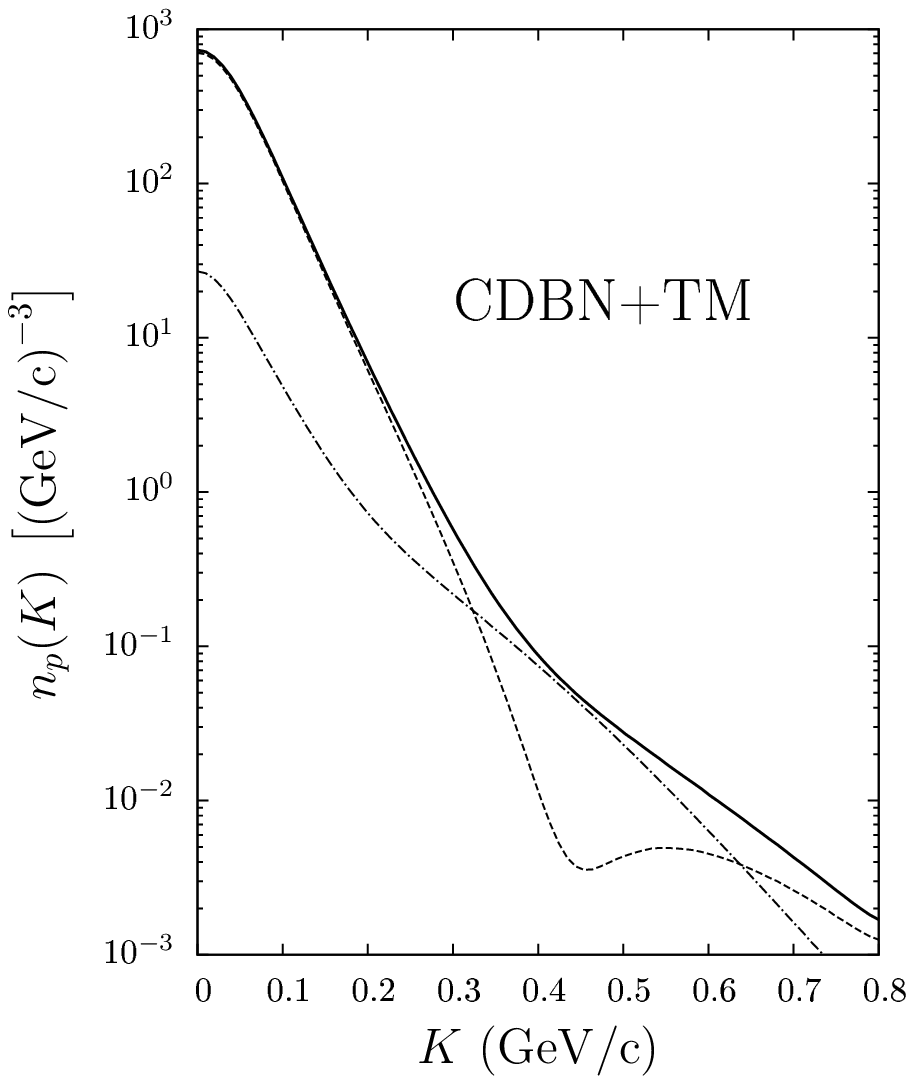}
 \vskip-3mm\caption{
Contributions of the main spin structures to the proton momentum distribution in $^3$He.
The notations are the same as those in Fig.~\ref{fig:neutMD_PW}.
\label{fig:protMD_PW}
 }
 \end{figure}
The momentum distribution of protons in $^3$He is given by
\be\label{p(k)_definition.1}
\begin{split}
n_p(K )=&\tfrac12\sum_\nu \int d^3p d^3q \left[\delta(\mathbf p_1 -
\mathbf K ) + \delta(\mathbf p_2 - \mathbf K)\right] \\
& \hspace{3.cm} \times
\left|\Psi_\nu (\mathbf p,\mathbf q)\right|^2,
\end{split}
\ee
where $\mathbf K$ is the proton momentum in $^3$He and the factor $\frac12$ comes
from averaging over the nucleus magnetic quantum numbers. Due to
identity of the protons, this expression is reduced to
\be\label{p(k).2}
\begin{split}
n_p(K ) = &
\sum_\nu \int d^3p d^3q \delta(\mathbf p_1 - \mathbf K )\left|\Psi_\nu (\mathbf p,\mathbf q)\right|^2 \\
\equiv &
\sum_\nu \int d^3p d^3q \delta(\mathbf p -\tfrac12\mathbf q - \mathbf K )
\left|\Psi_\nu (\mathbf p,\mathbf q)\right|^2 \\
 = & 2 \int d^3p d^3q \delta(\mathbf p -\tfrac12\mathbf q - \mathbf K)\rho_p(p,q,x)\\
 =&32\pi \int_0^\infty dp p^2 \int_{-1}^1 d\cos \theta_p\rho_p(p,q,x),
 \end{split}
\ee
where $\theta_p$ is angle between $\mathbf p$ and $\mathbf K$,
and
\be\label{p(k).2.a}
\begin{split}
\rho_p(p,q,x)= &
\left[C_{10}(p,q) + P_2(x)C_{12}(p,q) \right]^2 \\
 & \hspace{-1.3cm} +
 \left[P_1(x)C_{21}(p,q) + P_3(x)C_{23}(p,q) \right]^2\\
& \hspace{-1.3cm}
+
 P_1^2(x)\left[C_{61}^2(p,q) + C_{71}^2(p,q) \right]\\
 &\hspace{-1.3cm}+
 \tfrac9{10}\left(1 + \tfrac 13 x^2\right)
 \left[C_{80}(p,q) + P_2(x)C_{82}(p,q) \right]^2\\
 & \hspace{-1.3cm}+
 \left(-1 + 3 x^2\right)
 P_1^2(x)C_{61}(p,q) C_{71}(p,q)
 \\
& \hspace{-1.3cm}+
 \sqrt{\tfrac{24}5}x P_1(x)
 \left[C_{61}(p,q) + C_{71}(p,q)\right]\\
& \times
\left[C_{80}(p,q) + P_2(x)C_{82}(p,q) \right].
 \end{split}
\ee
In the final line of Eq.~(\ref{p(k).2}),
$q$ and $x$ are considered as functions of $K$,
$p$, and $\theta_p$.

Using $\mathbf K=\mathbf p -\frac12 \mathbf q$, we get
\be\label{p(k).3}
\begin{split}
 & q=2\sqrt{K^2 + p^2 -2Kp\cos\theta_p}\, ,\\
 & x=\frac{p-K\cos\theta_p}{\sqrt{K^2 + p^2 -2Kp\cos\theta_p}}.
 \end{split}
\ee
On the 2-dimensional grid, the integral over $dp$ can be reduced to the sum
$\sum_{i=1}^{n_p} w_i$, where $w_i$ is an element on the grid $(p,q,x)$. In turn, the integral
over $d\cos \theta_p$ becomes
\be\label{p(k).4}
\int_{-1}^1 d\cos \theta_p P_{\ell_1'}(x) P_\ell(x)C_{m\ell_1}(q,p_i) C_{m'\ell'}(q,p_i).
\ee
The variables $x$ and $q$ are defined by Eqs.~(\ref{p(k).3}),
therefore
$q$ cannot be on
the grid $(p,q,x)$. Nevertheless, the functions $C_{m\ell_1}(q,p_i)$
and $C_{m'\ell'}(q,p_i)$ at fixed  
$p_i$, $K$, and $\cos \theta_p$ can be obtained by
a linear interpolation from their values given on the grid $(q,p)$.

The
contribution of the spin structure 1 and sum of contributions coming from
the spin structures 2, 6, 7, and 8 to the momentum
distribution of protons in $^3$He are displayed in Fig.~\ref{fig:protMD_PW}.
\section{Empirical momentum distribution \label{sect:EMD}}
\begin{figure}
\includegraphics[
 height=0.375\textheight]{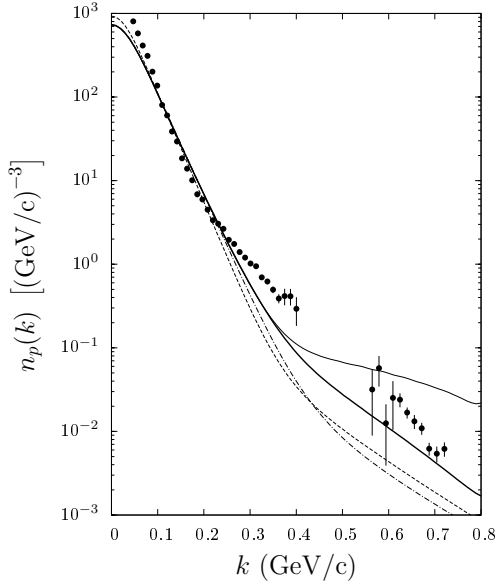}
\vskip-3mm\caption{Momentum distribution of protons in $^3$He calculated with
2N+3N interactions.
 The notations of the curves are the same as those in Fig.~\ref{fig:neutMD}.
Circles represent the empirical momentum distribution extracted
from the experimental data \cite{ALPHA}. Here,
$k$ is the LFD variable, as defined in Section~\ref{sect:EMD}.
\label{fig:protMD} }
\end{figure}%

Here, we compare the calculated proton momentum distributions with experimental results 
extracted from the $^{12}$C($^3$He$,p$) breakup cross-section measured at
$p_{^3\text{He}}=$~10.8~GeV/$c$  with the
emission of the proton-fragments at 0$^\circ$ \cite{ALPHA}.

\subsection{Empirical momentum distribution \label{sect:EMDdef}}

To compare the calculated momentum distribution with experiment, it is necessary to
establish a connection between the momentum {\bf K} (which is a theoretical quantity)
and the measured proton momentum.
In the non-relativistic case,  it is of a common use to postulate that
$\mathbf K = \mathbf k^\ast$, where $\mathbf k^\ast$
is the proton momentum is the $^3$He rest frame. But in the relativistic case,
as that of the experiment \cite{ALPHA}, it is incorrect.

The more adequate description has been suggested long time ago
within the so-called ``minimal relativization scheme''.
This approach was discussed in Ref.~\cite{KS} in detail. Therefore, we recall only
the main points here.

In the framework of this scheme, the momentum $\mathbf{K}$  is to be identified with the
``relativistic internal momentum'' $\mathbf{k}=(\mathbf{k}_\perp,k_{\|})$,
which appears in the dynamics on the light  front (LFD), instead of the
non-relativistic $\mathbf k^\ast$. The LFD is often called as the
``dynamics in the infinite momentum frame'' (IMF).
(The IMF is defined as a limiting reference frame, which is moving, with respect
to the laboratory frame, in the negative $z$-direction with a
velocity close to the speed of light.) In other words, it is the
$\mathbf k$ variable corresponding to the variable $\mathbf K$
used in the previous sections.
The important question is: ``In which way the light-front variable $\mathbf k$ is
related to the measured momentum of a $^3$He fragment?''

In the IMF dynamics, the wave function of a bound state is described in terms of two variables, $\alpha$ and
 $\mathbf k_\perp$. Let us consider $^3$He as a (proton+2N)
system with masses $m$ and $\mathcal M_{2N}$ respectively;
then $\alpha$ and $\mathbf k_\perp$ are
defined by
\be\label{IMF.1}
\alpha = \frac{E_p^{\mathrm{ lab}} + k_{\|}^{\mathrm{ lab}}}
              {E_{^3\mathrm{He}}^{\mathrm{lab}} + P_{\|}^{\mathrm{ lab}}}, \qquad
              \mathbf{k}_\perp = k_\perp^{\mathrm{ lab}},
\ee
where $p=(E_p^{\mathrm{ lab}},\mathbf k_\perp^{\mathrm{ lab}}, k_{\|}^{\mathrm{ lab}})$
 and $P = (E_{^3\mathrm{He}}^{\mathrm{ lab}}, \mathbf 0_\perp, P^{\mathrm{ lab}})$ are the proton
 and $^3$He 4-momenta in the laboratory frame. In terms of $\alpha$ and $\mathbf{k}_\perp$,
the effective mass squared of the ($p$+2N) system becomes
\be \label{IMF.2}
 \mathcal M^2_{p+2\mathrm N} = \frac{\alpha m^2 + (1 - \alpha)\mathcal M^2_{2\mathrm N}
 + \mathbf{k}_\perp^2}{\alpha(1-\alpha)}\, ,
\ee
and the longitudinal component of the $\mathbf{k}$ momentum is given by
\be\label{IMF.3}
k_{\|} = \pm\sqrt{\frac{\lambda(\mathcal M_{p+2\mathrm N}^2,
\mathcal M_{2\mathrm N}^2,m^2)}
{4\mathcal M_{p+2\mathrm N}^2}-\mathbf{k}_\perp^2},
\ee
where  $\lambda(a,b,c)=a^2+b^2+c^2-2ab-2ac-2bc$. In Ref.~\cite{KS},
it was argued that, because the mean momentum square in the pair
$\langle q^2\rangle \ll m^2$, one can take $\mathcal M_{2\mathrm N} \approx 2m$.

From the kinematical conditions of experiment \cite{ALPHA}, it follows
that $\mathbf q_\perp = 0$ and $\mathbf k_\perp = 0$.
In this case, the signs $"-"$ and $"+"$ are chosen for $\alpha<\frac13$ and
$\alpha>\frac13$, respectively; the IMF momentum $\mathbf k$ is reduced to the
momentum $\mathbf k^*$ for $\alpha \approx \frac13$.

The integral
\be\label{rel_p_number}
\begin{split}
& \int d^3k n_p(k) \\
& =
\int_0^1d\alpha\int d^2k_\perp \frac{\varepsilon_p(k)\varepsilon_{2N}(k)}{\alpha(1-\alpha)\mathcal M^2_{p+2\mathrm N}} n_p(k)
=2, \\
&
\varepsilon_p(k)=\sqrt{m^2+k^2}, \qquad \varepsilon_{2N}(k)=\sqrt{\mathcal M_{2\mathrm N}^2+k^2}
\end{split}
\ee
gives the number of protons in $^3$He, and the following expression can be considered
as the relativized momentum distribution of protons in $^3$He:
\be\label{rel_d_mom_dist}
n_p^{\mathrm{rel}}(\alpha,\mathbf k_\perp)= \frac{\varepsilon_p(k)\varepsilon_{2N}(k)}
{\alpha(1-\alpha)\mathcal M^2_{p+2\mathrm N}} n_p(k).
\ee

After that, in the framework of the IMF dynamics, the invariant differential
cross-section of the $A(^3\mathrm{He},p)$ breakup is given by
\be\label{CS(t,p)}
\begin{split}
& E_p\frac{d^3\sigma}{d\mathbf p_p}=f_\mathrm{kin}^{(p)} \sigma_d (1-\alpha) n_p^{\mathrm{rel}}(\alpha,\bf k_\perp),\\
& f_\mathrm{kin}^{(p)}=
\frac{\lambda^{\frac12}(W,\mathcal M_{2N}^2,M_A^2)}{2\alpha M_A P}\, ,
\end{split}
\ee
where $W$ and $M_A$ are the missing mass squared and the mass of the target nucleus,
respectively; the $\sigma_d$ factor plays the role of a normalization factors.

Equation~(\ref{CS(t,p)}) can be used to extract the proton momentum distribution in $^3$He.

It is clear that this equation was derived in the framework of the impulse approximation.
Nevertheless, one may expect that the momentum distribution extracted
from experimental data effectively includes effects beyond the impulse approximation, in particular, coming from the quark structure of $^3$He.
Therefore it was called in Ref.~\cite{KS} as ''empirical momentum distributions''
(EMD) of the protons in $^3$He.
\subsection{ Comparison with experiment \label{sec:EMDComparison}}
In Fig.~\ref{fig:protMD},
we compare results of our calculations for EMD extracted from
data \cite{ALPHA}, as well as with the calculations of Ref.~\cite{KS}, based on 2N
interactions only.

There is rather good agreement between calculations and EMD data at
$k \lesssim$~250~MeV/$c$. At very small $k$ ($\lesssim$~50~MeV/$c$) an enhancement of
EMD data over theoretical curves is as obvious for the $^3$He case 
as it was for the deuteron data.
This effect may be naturally explained as a result of contributions of the
Coulomb interaction to the breakup with the registration of a charged fragment at zero
emission angle. Note that a similar enhancement takes place also in EMD
of protons in a deuteron, extracted from data on the $^{12}$C($d,p$)
breakup \cite{JINR(dp)}.
The results of calculations published in Ref.~\cite{KKE} and based on the
Glauber-Sitenko model support the interpretation of this enhancement in
the momentum distribution of protons in a deuteron
as a manifestation of the Coulomb interaction. Of course, the
final state interaction also might be significant in the region of
small $k$.
In case of the deuteron breakup, this effect was (in part, at least)
taken into account in Ref.~\cite{KKE}.

From the comparison of our results with EMD data under discussion,
as well as with results published in Ref.~\cite{KS} at $k >$~250~MeV/$c$, the
following conclusions can be drawn:
\begin{itemize}
 \item There is rather visible qualitative disagreement between the
 calculations  and EMD of protons in $^3$He.
 \item Contribution of 3N interactions becomes significant in the
 $k >250$~MeV/$c$ region, but cannot explain completely the
disagreement between the data on EMD of protons and calculations based on 2N
interactions only.
\item  Version of the $^3$He wave function based on the CDBN+TM potential looks more
preferable than the version based on the AV18+U9 potential, because the latter
strongly overestimates the existing EMD data at very high momenta
(above 600 MeV/$c$).
\end{itemize}
\section{Conclusions\label{sec:concl}}
The momentum distributions of neutrons and protons in $^3$He have been calculated,
by using the so-called ``operator'' form for the
description of the 3N system. We used results of Ref.~\cite{Fachruddin_et_al}, where
the calculations of the necessary scalar functions (appearing in the operator form
representation of the bound 3N system) were performed with two potentials, which
involve the effective 3N interactions, 2N interaction AV18 \cite{AV18} with the
interaction Urbana-IX \cite{U9} (AV18+U9),  and 2N interaction CD-Bonn \cite{CDB} with
insertion of the 3N Tucson-Melbourne interaction~\cite{TM} (CDBN+TM).

We compare our results with calculations of Ref.~\cite{KS},
which do not take
the 3N interactions into account, and conclude that the
3N interactions become essential at
the large internal momentum $K>$250~MeV/$c$ of a nucleon in the bound 3N system.

We also compare the calculated momentum distribution of protons with the
so-called empirical momentum distribution
in $^3$He, extracted from the $(^3\mathrm{He},p)$ breakup cross-section \cite{JINR(dp)},
and conclude that 3N interactions reduce the
disagreement between theory and experiment at $k>$250~MeV/$c$.
Nevertheless, this disagreement does not completely disappear even in
the case where the 3N interactions are taken into account.

That means, that non-nucleonic degrees of freedom in $^3$He, as well as mechanisms
beyond the so-called ``\,impulse approximation'' become important in the $^3$He
breakup at $k>$250~MeV/$c$ and all other processes, where the nucleon-constituents
of this nucleus (as well as other nuclei) are very close (at distances $<$0.8~fm)
to each other.


\begin{thebibliography}{99}

\bibitem{Ermisch}
K.~Ermisch et al., Phys. Rev. C {\bf 71}, 064004 (2005).

\bibitem{Kamada}
H.~Kamada, O.~Shebeko, and A.~Arslanaliev, Few-Body Syst. {\bf 58}, 70 (2017).

\bibitem{DuSh}
I.~Dubovyk and O.~Shebeko, Few Body Syst. {\bf 48}, 109 (2010).

\bibitem{ALPHA}
V.G.~Ableev et al., JETP Lett. {\bf 45}, 596 (1987); Pis'ma v ZhETP {\bf 45}, 467 (1987).

\bibitem{Baru_et_al}
V.~Baru, J.~Haidenbauer, C.~Hanhart, J.A.~Niskanen, Eur. Phys. J. {\bf A 16}, 437 (2003).

\bibitem{KotlyarShebeko}
V.V.~Kotlyar and A.V.~Shebeko, Zeitschrift f\"{u}r Physik A {\bf 327}, 301 (1987).

\bibitem{KotlyarShcheglova}
V.V.~Kotlyar and A.A.~Shcheglova,
Vist. Khark. Univ. $N^{\circ}$~832, Ser. Fiz. ``Yad., Chast., Polya'', Issue 4 (40), 11 (2008).

\bibitem{KotlyarJourdan}
V.~Kotlyar and J.~Jourdan, Problems of Atomic Science and Technology. Series: Nucl. Phys. Investigations {\bf 6(45)}, 24 (2005).

\bibitem{GerjouySchwinger}
E.~Gerjuoy and J.~Schwinger, Phys. Rev. {\bf 61}, 138 (1942).

\bibitem{RaritaSchwinger}
W.~Rarita and J.~Schwinger, Phys. Rev. {\bf 59}, 436 (1941); Phys. Rev. {\bf 59}, 556 (1941).

\bibitem{SDG}
I.V.~Simenog, I.S.~Dotsenko, and B.E.~Grinyuk, Ukr.~J.~Phys. {\bf 47}, 129 (2002).

\bibitem{DS}
I.S.~Dotsenko and I.V.~Simenog, Ukr.~J.~Phys. {\bf 51}, 841 (2006).

\bibitem{Fachruddin_et_al}
I.~Fachruddin, W.~Gl\"ockle, Ch.~Elster, and A.~Nogga, Phys. Rev. C {\bf 69}, 064002 (2004).

\bibitem{AV18}
R.B.~Wiringa, V.G.J.~Stoks, and R.~Schiavilla, Phys. Rev. C {\bf 51}, 38 (1995).

\bibitem{U9}
B.S.~Pudliner et al., Phys. Rev. C {\bf 56}, 1720 (1997).

\bibitem{CDB}
R.~Machleidt, Phys. Rev.  C {\bf 63}, 024001 (2001).

\bibitem{TM}
S.A.~Coon et al.,  Nucl. Phys. A {\bf 317}, 242 (1979);
S.A.~Coon and W.~Gl\"ockle,  Phys. Rev. C {\bf 23}, 1790 (1981).

\bibitem{grid}
\url{http://www.phy.ohiou.edu/~elster/h3wave }

\bibitem{KS}
A.P.~Kobushkin and E.A.~Strokovsky, Phys. Rev. C {\bf 87}, 024002 (2013).

\bibitem{SPW}
R.~Schiavilla, V.R.~Pandharipande, and R.B.~Wiringa, Nucl. Phys. A {\bf 449}, 219 (1986).

\bibitem{Paris}
M.~Lacombe et al., Phys. Rev. C {\bf 21}, 861 (1980).

\bibitem{JINR(dp)}
V.G.~Ableev et al., Nucl. Phys. A {\bf 393}, 491 (1983) and A {\bf 411}, 541 (E) (1983);
V.G.~Ableev et al., JINR Rapid Comm. 1[52]-92 10 (1992);
V.G.~Ableev et al., JINR Rapid Comm. 1[52]-92 5 (1992) and Yad. Fiz. {\bf 37}, 132 (1983).

\bibitem{KKE}
A.P.~Kobushkin and Ya.D.~Krivenko-Emetov, Ukr.~J.~Phys. {\bf 53}, 751 (2008).

\end{thebibliography}
\end{document}